# Microwave Magnetoelectric Effects in Single Crystal Bilayers of Yttrium Iron Garnet and Lead Magnesium Niobate-Lead Titanate

S. Shastry and G. Srinivasan

*Physics Department, Oakland University, Rochester, MI 48309*

M. I.Bichurin, V. M. Petrov and A.S.Tatarenko

*Department of Physics and Engineering, Novgorod State University,*

*B. S. Peterburgskaya St. 41, 173003 Veliky Novgorod, Russia*

ABSTRACT

The first observation of microwave magnetoelectric (ME) interactions through ferromagnetic resonance (FMR) in bilayers of single crystal ferromagnetic-piezoelectric oxides and a theoretical model for the effect are presented. An electric field $E$ produces a mechanical deformation in the piezoelectric phase, resulting in a shift $\delta H_E$ in the resonance field for the ferromagnet. The strength of ME coupling is obtained from data on $\delta H_E$ vs $E$. Studies were performed at 9.3 GHz on bilayers of (111) yttrium iron garnet (YIG) films and (001) lead magnesium niobate-lead titanate (PMN-PT). The samples were positioned *outside* a $TE_{102}$-reflection type cavity. Resonance profiles were obtained for $E$ = 0-8 kV/cm for both in-plane and out-of-plane magnetic fields $H$. Important results are as follows. (i) The ME coupling in the bilayers is an order of magnitude stronger than in polycrystalline composites and is in the range 1-5.4 Oe cm/kOe, depending on the YIG film thickness. (ii) The coupling strength is dependent on the magnetic field orientation and is higher for out-of-plane $H$ than for in-plane $H$. (iii) Estimated ME constant and its dependence on volume ratio for the two phases are in good agreement with the data.

## I. INTRODUCTION

The electromagnetic coupling in a ferromagnetic-ferroelectric heterostructure is facilitated by mechanical deformation.[1-7] In an applied magnetic field, for example, deformation arises due to magnetostriction and results in an induced polarization because of piezoelectric effect. The induced polarization $P$ is related to the magnetic field $H$ by the expression, $P = \alpha H$, where $\alpha$ is the second rank magnetoelectric (ME) susceptibility tensor. We are interested in the dynamic ME effect; for an ac magnetic field $\delta H$ applied to a biased sample, one measures the induced electric field $\delta E$. The ME voltage coefficient $\alpha_E = \delta E/\delta H$ and $\alpha = \varepsilon_o \varepsilon_r \alpha_E$ where $\varepsilon_r$ is the relative permittivity. Our recent studies on bilayers and multilayers of ferrite-lead zirconate titanate (PZT) and manganite-PZT indicated a giant ME effect with $\alpha_E$ as high as 1500 mV/cm Oe at low frequencies.[5-7]

The frequency $f$ dependence of $\alpha_E$ is of interest. Investigations so far have mainly focused on low frequency (10 Hz – 500 kHz) ME phenomena. Our measurements in ferrite-PZT indicate a general increase in $\alpha_E$ with $f$ over the range 10 Hz – 10 kHz.[5] With further increase in frequency, a dramatic increase in $\alpha_E$ is observed at electromechanical resonance (EMR) due to radial acoustic modes.[8] The EMR occurs at 100-400 kHz for disk shaped samples of diameter 25-10 mm. One anticipates a similar resonance in $\alpha_E$ vs $f$ at a higher frequency due to thickness modes in the samples.

This work constitutes the first investigation on the nature of ME interactions at microwave frequencies (9-10 GHz) in single crystal bilayers of ferrite/ferroelectrics. Ferromagnetic resonance (FMR) is a powerful tool for studies. An electric field E applied to the composite produces a mechanical deformation in the ferroelectric that in turn is coupled to the ferrite, resulting in a shift $\delta H_E$ in the resonance field. Information on the nature of ME coupling could therefore be obtained from data on shift $\delta H_E$ vs $E$.[9-13] Such measurements of course require composites with low line width ferromagnetic phases such as yttrium iron garnet (YIG). Studies on *bulk polycrystalline samples* of 90 wt.% YIG -10% PZT revealed weak ME interactions because of low PZT concentration.[9] But when the concentration of PZT was increased, FMR line broadening masked electric field effects on the resonance field. Such line broadening problems, however, could easily be eliminated in a *layered* structure of YIG-PZT. A further reduction in the line width is possible with the use of *single crystal* YIG in the layered samples.

Here we present results of investigations on microwave ME interactions in single crystal bilayers of YIG and lead magnesium niobate-lead titanate (PMN-PT). In addition to the advantage of low line widths for accurate determination of ME coupling, our theory predicts an order of magnitude enhancement in ME coupling in single crystals compared to polycrystalline samples.[10,11]

Further, an understanding of the phenomenon requires the use of single crystals with known crystallographic, material and interface parameters. The measurements at 9.3 GHz indicate strong ME coupling in the bilayers. The ME coupling is found to be dependent critically on the H orientation and YIG film thickness. We also extend our earlier theoretical model for microwave ME coupling to include single crystal bilayers. Expressions have been obtained for resonance field shifts in terms of ME constants. Estimated field shifts are in very good agreement with data.

## II. EXPERIMENT

Bilayers (4 mm x 4 mm) were fabricated with 1-110 μm thick epitaxial (111) YIG films on gadolinium gallium garnet (GGG) substrates and 0.5 mm thick (001) PMN-PT for the piezoelectric phase.[14] Gold electrodes were deposited on PMN-PT for electrical contacts. A thin layer (<0.08 mm) of epoxy was used to bond YIG to PMN-PT. The YIG film on the non-contact side was removed and GGG thickness was reduced to 0.1 mm by polishing.

Microwave ME measurements at 9.3 GHz were performed using a modified Varian electron spin resonance spectrometer. A frequency stabilized klystron and a $TE_{102}$ reflection type cavity with $Q$ of 2000 were used. The incident power on the cavity was kept small to about to 0.1 mW, corresponding to an rf magnetic field of 1.7 mOe, to avoid any sample heating at resonance. Holes (1 mm diameter) were made at the center of the cavity bottom and at λ/4 from the bottom on the narrow side. The holes were necessary for measurements with the sample *outside* the cavity in order to eliminate cavity overloading during resonance absorption. The YIG films were first characterized by magnetization and FMR. The resonance field $H_r$ for $H$ parallel or perpendicular to (111) plane was in agreement with expected values for the effective saturation induction $4\pi M_{eff} = 4\pi M_s + H_a$ = 1.8-1.9 kG, where $4\pi M_s$ is the saturation magnetization and $H_a$ is the anisotropy field. We also found evidence for anisotropy in the (111) plane, possibly due to strain at the YIG-GGG interface. The YIG/PMN-PT bilayer was then mounted just outside the cavity holes for ME studies. Measurements were done with $E$ perpendicular to the bilayer plane and for the following $H$ field orientations. (i) $H$ parallel to <111> of YIG: for this case the sample was mounted on the narrow side. (ii) $H$ parallel to (111) of YIG: bilayer was mounted at the cavity bottom. Absorption vs $H$ profiles were recorded for $E$ = 0-8 kV/cm.

## III. RESULTS

Figure 1 shows representative data on electric field effects on FMR at 9.3 GHz in a bilayer of 13.2 μm YIG and PMN-PT. The sample was placed outside the hole on the cavity narrow side for these data. Both E and H were along <111> of YIG. For $E$ = 0, the figure shows FMR at $H_r$ = 5235 Oe with a line width $\Delta H$ = 6±2 Oe (vs 50-100 Oe for polycrystalline YIG).[15] Some magnetostatic modes are also seen in the microwave absorption vs $H$ profile. The condition for FMR when E = 0 is given by

$$\omega/\gamma = H_r - 4\pi M_{eff} \quad (1)$$

where $\omega$ is the angular frequency and $\gamma$ is the gyromagnetic ratio (and $4\pi M_{eff}$ is defined earlier). For the above field and frequency and $\gamma$ = 2.8 GHz/kOe, one estimates $4\pi M_{eff}$ =1.91 kG which is agreement with expected values for YIG.[15] With the application of $E$ = 1 kV/cm, we notice in Fig. 1 an up-shift in $H_r$ by $\delta H_E$ = 4 Oe, but $\Delta H$ remains the same. Further increase in $E$ results in an increase in $H_r$. Although the data in Fig.1 do not indicate any broadening or narrowing of FMR, the intensity of the absorption increases with E. The magnetostatic modes, however, broaden and merge as $E$ is increased. The field shift data in Fig. 1 cannot be due to sample heating. Based on the resonance condition in Eq. (1) any sample heating is expected to *decrease* the sample magnetization and thus *decrease* the external $H_r$ necessary for resonance. But the ME coupling manifests as an *increase* in $H_r$.

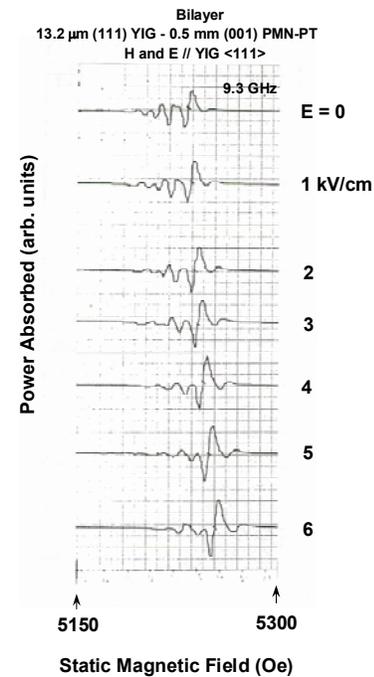

Fig. 1: Ferromagnetic resonance (FMR) at 9.3 GHz in a bilayer of (111) yttrium iron garnet (YIG) on gadolinium gallium garnet (GGG) substrate and (001) lead magnesium niobate-lead titanate (PMN-PT). A 13.2 micron thick YIG film and 0.5 mm thick PMN-PT were used. The profiles of absorption vs static magnetic field $H$ are shown for a series of electric field $E$ applied to the bilayer and for $E$ and $H$ applied perpendicular to the bilayer plane and parallel to <111> axis of YIG. The profiles show an increase in the resonance field with $E$.

Similar measurements were made on bilayers with a series of YIG thickness and the results are shown in Fig.2. The data on $\delta H_E$ vs $E$ are for $H$ and $E$ along <111> of YIG as in Fig. 1. Consider first the results for the bilayer with 4.9 μm YIG. A linear increase in $\delta H_E$ with $E$ is indicative of the absence of bilinear ME effects. The ME constant $A = \delta H_E/E$ obtained from the data is 5.4 Oe cm/kV. Upon increasing the YIG thickness to 13.2 μm, one finds a similar behavior as for the bilayer with 4.9 μm film, but the ME constant is reduced to 4.4 Oe cm/kV. A further reduction in $A$ to 2.3 Oe cm/kV is measured for the bilayer with 110 μm thick YIG.

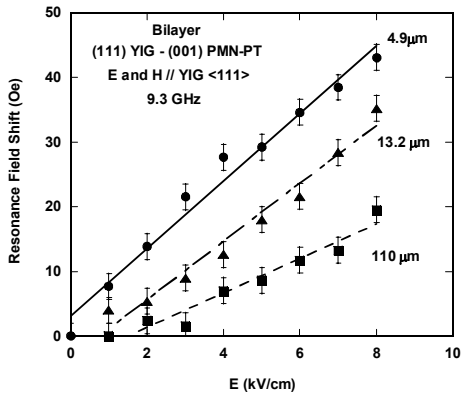

Fig. 2: Microwave magnetoelectric effect at 9.3 GHz in bilayers of (111) YIG-(001) PMN-PT. The shift in the resonance field $\delta H_E$, measured from profiles as in Fig.1, is shown as a function of $E$ for a series of YIG film thickness. The lines are linear fit to the data.

Ferromagnetic resonance studies were also performed for $H$ parallel to (111) of YIG by placing the bilayer outside the hole at the center of cavity bottom. Data showed an increase in the resonance field with the application of an electric field to PMN-PT. But the magnitude of $\delta H_E$ and the coupling constant $A$ were relatively small compared to values for out-of-plane $H$. For the bilayer with 110 μm thick YIG, for example, $A$ decreased from 2.3 Oe cm/kV for out-of-plane $H$ to 1 Oe cm/kV for in-plane $H$.

Since the microwave ME effect arises due to piezoelectric effect in PMN-PT, $\delta H_E$ and $A$ are very much dependent on the volumes of YIG and PMN-PT.[10] Figure 3 shows the measured variation of $\delta H_E$ for $E$ = 8 kV/cm with the volume ratio for both in-plane and out-of-plane $H$. The field shift (and the ME constant $A$) decrease with increasing volume of YIG as predicted in our model.[10]

Next we compare the ME constant for the single crystal bilayers with $A$ for similar ferrite-piezoelectric bulk and layered samples. For bulk YIG-PZT composites, the measured $A$ = 0.33 Oe cm/kV is an order of magnitude smaller than current values for single crystal bilayers.[9] We recently performed microwave ME studies on multilayers of LFO-PZT.[13] The sample contained 16 layers of LFO and 15 layers of PZT, with a layer thickness of 15 μm. Data on the shift $\delta H_E$ with $E$ showed a linear dependence, indicative of the absence of measurable bilinear ME effects. From the data, an ME coefficient $A$ = 0.25 Oe cm/kV was obtained, in agreement with our theoretical estimates in Ref.10. Thus bulk YIG-PZT and layered LFO-PZT samples show an order of magnitude weaker ME coupling than for single crystal YIG/PMN-PT.

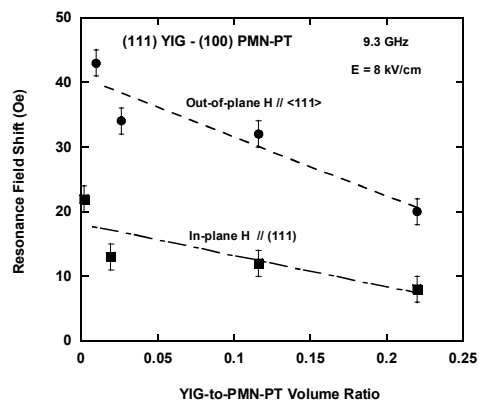

Fig. 3: Measured field shift at $E$ = 8 kV/cm as a function of YIG-to-PMN-PT volume ratio for in-plane and out-of-plane $H$. The lines are guide to the eye.

### IV. THEORY OF ME COUPLING AT FMR IN BILAYERS

We consider a two-layer structure consisting of a single crystal film YIG with cubic ($m3m$) symmetry and a poled single crystal plate PMN-PT with $\infty m$ symmetry about the poling axis. The influence of the electric field upon the piezoelectric phase may be described as:[10,11]

$$^pS_{ij} = {}^ps_{ijkl}{}^pT_{kl} - {}^pd_{kij}E_k. \qquad (2)$$

where $E_k$ is the electrical field intensity; $^pT_{ij}$, $^pS_{kl}$, $^pd_{kij}$, and $^pS_{ijkl}$ are the stress, strain, piezoelectric coefficient, and compliance tensors of the piezoelectric phase, respectively. Assuming the electric field to be directed along the poling axis, i.e. $E_3 = E$, $E_1 = E_2 = 0$ and taking into account two-index notation for tensors we get:

$$^pS_3 = {^pd_{33}}E + {^pS_{13}}(^pT_1 + {^pT_2}) + {^pS_{33}}{^pT_3},$$
$$^pS_1 = {^pd_{31}}E + {^pS_{11}}{^pT_1} + {^pS_{12}}{^pT_2} + {^pS_{13}}{^pT_3},$$
$$^pS_2 = {^pd_{31}}E + {^pS_{12}}{^pT_1} + {^pS_{11}}{^pT_2} + {^pS_{13}}{^pT_3}.$$
(3)

We assume that the poling axis of the piezoelectric phase coincides with <111> axis of the magnetostrictive phase. In that case, the compliance tensor of the magnetostrictive phase has the form:

$$^mS_{ijkl} = {^mS_{i'j'k'l'}}\beta_{ii'}\beta_{jj'}\beta_{kk'}\beta_{kk'} \qquad (4)$$

where $(1, 2, 3)$ is a coordinate system in which the axis 3 is directed along the equilibrium magnetization and $\beta$ is matrix of direction cosines of axes $(1, 2, 3)$ relative to the crystallographic coordinate system $(1', 2', 3')$. For the magnetostrictive phase we get:

$$^mS_3 = {^mS_{13}}(^mT_1 + {^mT_2}) + {^mS_{33}}{^mT_3}, \qquad (5)$$
$$^mS_1 = {^mS_{11}}{^mT_1} + {^mS_{12}}{^mT_2} + {^mS_{13}}{^mT_3},$$
$$^mS_2 = {^mS_{12}}{^mT_1} + {^mS_{11}}{^mT_2} + {^mS_{13}}{^mT_3}.$$

Here $^mT_i$ and $^mS_k$ are the stress and strain tensors of the magnetostrictive phase, respectively. Next we consider the unclamped bilayer sample of YIG – PMN-PT composite with the following boundary conditions

$$^pS_1 = {^mS_1}, \quad {^pS_2} = {^mS_2}, \quad {^pT_3} = {^mT_3} = 0,$$
$$^pT_1 = -{^mv}/{^pv} \cdot {^mT_1}, \quad {^pT_2} = -{^mv}/{^pv} \cdot {^mT_2}, \qquad (6)$$

where $^mv$ and $^pv$ are the volume fraction of magnetostrictive phase and piezoelectric phase, respectively. Solution of Eqs.(3) and (5) taking into consideration Eq. (6) yields:

$$^mT_1 = {^mT_2} = -{^pd_{31}} E {^pv}/[{^pv}({^mS_{11}} - {^mS_{12}}) + {^mv}({^pS_{11}} - {^pS_{12}})] \qquad (7)$$

The influence of an external constant electric field $E$ upon a magnetic resonance spectrum can be described by means of an additional term in the thermodynamic potential that can be found by taking into account elastic, magnetoelastic, magnetostrictive, piezoelectric contributions under certain specified boundary conditions. The additional energy term in our present case is represented by:

$$\Delta W_{ME} = -3[\lambda_{100}(T_1 \cdot M_1^2 + T_2 \cdot M_2^2 + T_3 \cdot M_3^2) + 2\lambda_{111}(T_6 \cdot M_1 \cdot M_2 + T_5 \cdot M_1 \cdot M_3 + T_4 \cdot M_2 \cdot M_3)]/2M_0^2.$$
(8)

Here the stress component $T_{i'j'}$ can be found from

$$T_{i'j'} = {^mT_{ij}}\beta_{ii'}\beta_{jj'}, \qquad (9)$$

where $^mT_{ij}$ are determined by Eq. (7).
The magnetic resonance condition has the well-known form:

$$\omega = \gamma \left\{ \left[ H_3 + \sum_i (N_{11}^i - N_{33}^i) M_0 \right] \left[ H_3 + \sum_i (N_{22}^i - N_{33}^i) M_0 \right] - \left( \sum_i N_{12}^i M_0 \right)^2 \right\}^{1/2},$$
(10)

where $H_3$ is the projection of an external magnetic field towards the equilibrium orientation of the magnetization and $N_{kn}^i$ are demagnetization factors describing shape anisotropy, crystalline anisotropy and due to ME interactions. Effective demagnetization factors due to the magnetic crystalline anisotropy and ME interaction are determined by following expressions:

$N^a_{11} - N^a_{33} = [2(\beta^4_{31'} + \beta^4_{32'} + \beta^4_{33'}) - 6(\beta^2_{11'}\beta^2_{31'} + \beta^2_{12'}\beta^2_{32'} + \beta^2_{13'}\beta^2_{33'})]H_a/M_0;$
$N^a_{22} - N^a_{33} = [2(\beta^4_{31'} + \beta^4_{32'} + \beta^4_{33'}) - 6(\beta^2_{21'}\beta^2_{31'} + \beta^2_{22'}\beta^2_{32'} + \beta^2_{23'}\beta^2_{33'})]H_a/M_0.$
$N^E_{kn} = 2b_{ijkn}T_{ij}\beta_{kk'}\beta_{nn'},$

where $b_{1111} = b_{2222} = b_{3333} = 3\lambda_{100}/(2M_0^2)$, $b_{1212} = b_{1313} = b_{2323} = 3\lambda_{111}/M_0^2$.
For magnetization directions considered here $N^a_{12} = 0$. Using Eq. (10) it is easily shown that the resonance line shift under the influence of the electric field to the first order in $N^E_{kl}$ has the form:

$$\delta H_E = -\frac{M_0}{Q_1}\left[Q_2\left(N_{11}^E - N_{33}^E\right) + Q_3\left(N_{22}^E - N_{33}^E\right) - Q_4 N_{12}^E\right],$$
(12)

where

$$Q_1 = 2H_3 + M_0 \sum_{i \neq E}\left[\left(N_{11}^i - N_{33}^i\right) + \left(N_{22}^i - N_{33}^i\right)\right];$$

$$Q_2 = \left[H_3 + M_0 \sum_{i \neq E}\left(N_{22}^i - N_{33}^i\right)\right];$$

$$Q_3 = \left[H_3 + M_0 \sum_{i \neq E}\left(N_{11}^i - N_{33}^i\right)\right];$$

$$Q_4 = 2M_0 \sum_{i \neq E} N_{12}^i.$$

Here $H_3$ is determined by Eq. (10).

Next we consider two important cases of magnetic field and stress directions for unclamped bilayers: (i) H along <111> and (ii) H parallel to (111). For the first case, the matrix $\beta$ has the form:

$$\beta_{ij'} = \begin{bmatrix} \sqrt{6}/3 & \sqrt{3}/3 & \sqrt{3}/3 \\ -\sqrt{6}/6 & \sqrt{2}/2 & \sqrt{3}/3 \\ -\sqrt{6}/6 & -\sqrt{2}/2 & \sqrt{3}/3 \end{bmatrix}$$

and the geometrical demagnetization factors are
$N^F_{11} = N^F_{22} = 0$, $N^F_{33} = 4\pi$.

For the second case of $H$ parallel to (111) plane and to <011>, the matrix $\beta$ has the form:

$$\beta_{ij'} = \begin{bmatrix} \sqrt{6}/3 & \sqrt{3}/3 & 0 \\ -\sqrt{6}/6 & \sqrt{3}/3 & -\sqrt{2}/2 \\ -\sqrt{6}/6 & \sqrt{3}/3 & \sqrt{2}/2 \end{bmatrix}.$$

and the geometrical demagnetization factors are

$N^F_{11} = N^F_{33} = 0$, $N^F_{22} = 4\pi$.

Resonance line shift under the influence of the electric field can be found from the equations provided here. Analytical expression for resonance line shift is too tedious; it is more practical to solve these equations numerically and to present it graphically.

## V. DISCUSSION

The theory in Sec.IV facilitates the estimation of field shifts for comparison with data. The shifts are dependent on the boundary conditions (clamped or unclamped bilayers), volumes of YIG, GGG and PMN-PT. The implied assumption here is that the mechanical deformation at the YIG/PMN-PT interface is transmitted uniformly throughout the YIG film and the GGG substrate. In general, the electric field induced shift of resonance line can be described using an ME constant A:

$$\delta H_E = A E_3 , \qquad (17)$$

and $A$ equals the resonance line shift for $E = 1\ kV/$cm. The model predicts a strong microwave ME interaction when (i) the volume fraction of the piezoelectric phase is sufficiently high, (ii) the piezoelectric component has a large piezoelectric coupling coefficient and (iii) the magnetic phase has a small saturation magnetization and high magnetostriction.

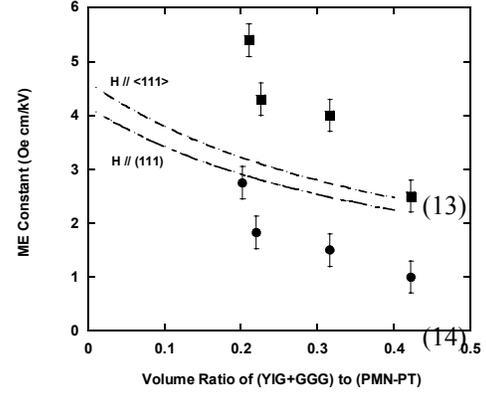

Fig. 4: Comparison of theory (lines) and data on the ME constant $A$ for in-plane (circles) and out-of-plane (squares) $H$. The $A$-values are shown as a function of ratio of volumes of (YIG+GGG) and PMN-PT.

The field shifts were calculated for by solving the above equations numerically for (i) $H$ parallel to <111> of YIG and (ii) $H$ parallel to <0$\bar{1}$1> in (111) plane of YIG. Figure 4 shows the estimates on the dependence of ME constant $A$ on the volume ratio of YIG+GGG to PMN-PT. The results are for PMN-PT and GGG layer thickness of 0.5 mm and 0.1 mm, respectively, and for the following material parameters.

<u>PMN-PT</u>: piezoelectric coupling constants $^p d_{31} = -600 \cdot 10^{-12}$ m/V, $^p d_{33} = 1500 \cdot 10^{-12}$ m/V and compliance constants $^p s_{11} = 23 \cdot 10^{-12}$ m$^2$/N; $^p s_{12} = -8.3 \cdot 10^{-12}$ m$^2$/N.

<u>YIG</u>: magnetostrictions $\lambda_{100} = -1.4 \cdot 10^{-6}$; $\lambda_{111} = -2.4 \cdot 10^{-6}$; $4\pi M_0 = 1750$ G; anisotropy field $H_a = -42$ Oe; and compliance constants $^m s_{11} = 4.8 \cdot 10^{-12}$ m$^2$/N; $^m s_{12} = -1.4 \cdot 10^{-12}$ m$^2$/N.

The figure shows theoretical values of $A$ vs. the volume ratio for both in-plane and out-of-plane $H$. One notices a decrease in $A$ with increasing YIG (or decreasing PMN-PT) volume. An important inference from Fig.4 is the prediction of a stronger ME coupling for out-of-plane $H$ than for in-plane fields and is due to higher magnetostriction along <111>. $A$ values determined from FMR data are compared with theory in Fig.4. The measured values are (i) higher than theoretical estimates for H //<111>, (ii) but are smaller when H // (111). There is overall good qualitative and quantitative agreement between theory and data.

The work described here is also of technological importance. The results in Figs.1-4 also open up the possibility for a new class of magnetoelectric signal processing devices, such as an FMR-based phase shifters and filters. The unique and novel feature in such devices is the tunability with an electric field. Most ferrite-based devices use a permanent magnet for the bias field and magnetic tuning of the devices even over a narrow frequency range is rather slow and is associated with large power consumption. It is clear from Fig.1 that with a ME composite, however, tuning could be accomplished with an easy to generate electric field. For example, a device based on YIG/PMN-PT operating at 9.3 GHz could be tuned over a frequency-width of 122 MHz (or $\delta H_r$ = 44 Oe) by an applied $E$=0-8 kV/cm. Profiles as in Fig.1 are useful for the design and characterization such devices.

## VI. CONCLUSION

Ferromagnetic resonance at x-band has been performed on single crystal (111)YIG/(001)PMN-PT bilayers to obtain information on the nature of ME interactions. The strength of the ME coupling has been estimated from the resonance field shift due to an applied electric field. Studies on bilayers with a series of thickness for the YIG film show an increase in the field shift with decreasing film thickness. The ME coupling is stronger for H perpendicular to the bilayer than for in-plane H. Estimates of resonance field shift with the YIG film thickness, volume ratio for the two phases and H-orientation are in agreement with the data.

**Acknowledgment**

The work at Oakland University was supported by grants from the National Science Foundation (DMR-0302254), the Army Research Office, and the Delphi Automotive Corporation. The work at Novgorod State University was supported by a grant from the Russian Ministry of Education (E02-3.4-278).